\documentclass[aps,prl,showpacs,10pt,superscriptaddress,
twocolumn,
]{revtex4-1}
\usepackage{amssymb,amsmath,amstext}                
\usepackage{graphicx}                                               
\usepackage{epstopdf}                                               
\usepackage{color}                                                     
\usepackage{bm}                                                        
\usepackage{appendix}                                              
\usepackage[utf8]{inputenc}
\usepackage{bbold}
\usepackage{float}
\usepackage[colorinlistoftodos]{todonotes}
\usepackage{latexsym}
\usepackage[colorlinks=true,citecolor=blue,linkcolor=magenta]{hyperref}

\begin{document}
\title{Rescattering effects in streaking experiments of strong-field ionization}

\author{Michał Mandrysz}
\email{michal.mandrysz@student.uj.edu.pl}
\affiliation{Instytut Fizyki im. Mariana Smoluchowskiego, Uniwersytet Jagiello\'nski,  \L{}ojasiewicza 11, 30-348 Krak\'ow, Poland}
\author{Matthias K\"ubel}
\affiliation{Department of Physics, Ludwig-Maximilians-Universit\"at Munich, D-85748 Garching, Germany}
\affiliation{Joint Attosecond Science Laboratory, National Research Council and University of Ottawa, 100 Sussex Drive, Ottawa, Ontario K1A 0R6, Canada}
\affiliation{Institute for Optics and Quantum Electronics, University of Jena, D-07743 Jena, Germany}
\author{Jakub Zakrzewski}
\affiliation{Instytut Fizyki im. Mariana Smoluchowskiego, Uniwersytet Jagiello\'nski,  \L{}ojasiewicza 11, 30-348 Krak\'ow, Poland}
\author{Jakub S. Prauzner-Bechcicki}
\affiliation{Instytut Fizyki im. Mariana Smoluchowskiego, Uniwersytet Jagiello\'nski,  \L{}ojasiewicza 11, 30-348 Krak\'ow, Poland}

\begin{abstract}
Strong field ionization provides a unique mean to address complex dynamics of an electron in competing Coulomb and laser fields.
Recent streak camera experiment (K\"ubel, et al., \cite{Kubel17}) analyzed asymmetries in the low-energy region of ejected electron momenta distribution and associated them with multiple rescattering of the electron on the parent ion. In this work we directly confirm the multiple rescattering nature of the asymmetry in the low-energy region. Such electron-ion collisions cannot be described within one-dimensional simulations even taking into account focal-volume averaging. Using time-dependent Schr\"odinger equation simulation in two dimensions supplemented by the insight from strong field semi-classical approximation we identify the dominant interference features of the complex photoelectron momentum maps and find their traces in the experiment. In particular, the holographic structures remain visible in experimental results averaged over carrier-envelope phase (CEP). In the case of individual momentum maps when CEP or delay between the two pulses is varied, the structures arising due to rescattering events are influenced by interfering electrons ionized at main peaks of the electric field. With an increase of experimental resolution, such structures record the electron dynamics on a sublaser-cycle time scale. 
\end{abstract}


\maketitle

\section{Introduction}
Interaction of atoms and molecules with intense, very short laser pulses results in many interesting phenomena, such as high-order harmonic generation (HHG), above-threshold ionization (ATI) or non-sequential multiple ionization (NSI). All these phenomena, are studied carefully since '80s of the last century with the use of both theoretical and experimental tools \cite{Becker12, agostini2004physics, krausz2009attosecond}. The experiments are becoming more and more refined, even allowing one to resolve dynamics of electron wave packets at the attosecond time-scale \cite{Kubel17,Ossiander17}. Such a situation sets the bar high for the theory. Theoretical description in many cases requires a non-perturbative treatment and, eventually, ends with simplified modeling and computer simulations~\cite{Efimov18}. This happens because, despite of enormous increase in available computer resources, full {\it ab initio} quantum calculations of processes involving more than one electron are typically beyond reach.

In view of the above, the Single Active Electron (SAE) approximation appears as a very powerful tool. In fact, one may relatively easily solve a full three-dimensional (3D) time-dependent Schr\"odinger equation (TDSE) for an atom with a single electron exposed to an external field with a given set of parameters \cite{Mosert16} without the necessity of referring to restricted geometry models \cite{Efimov18}. Importantly, the external field may have parameters, i.e., amplitude, frequency, envelope, carrier-envelope phase (CEP) and often duration, in ranges that are used in experiments. Already for two-electron atoms similar calculations are very demanding with respect to computer resources. To the best of our knowledge it was done only for helium \cite{Dundas99,Taylor03,Parker00,Emma11,Feist08,Pazourek12,Ossiander17}. The proper comparison of numerical results from such full 3D calculations with experimental data imposes the usage of various averaging techniques like focal volume or the Gouy phase averaging \cite{hoff2017tracing}. The averaging, in turn, calls for large data sets covering specific ranges of parameters resulting
, in the course of time,
in excessive computational demands. Thus, judiciously chosen restricted geometry model may be a good trade-off offering better agreement with experimental data at the lesser expense and granting enhanced insight. 
Indeed, numerical solutions of TDSE in various simplified models lead to results
closely resembling experimental data especially for some quantities of interest, such as dipole acceleration \cite{ciappina2012high}. 

In general, the interpretation of TDSE results requires the use of certain analytic methods such as the strong field approximation (SFA) \cite{Lewenstein94,amini2018symphony,peng2015tracing}. When applied to photoelectron momentum maps it becomes possible to separate the individual processes and trajectories that correspond to them. These trajectories interfere with each other and, ultimately, lead to a very complex final image. The reasoning can also be applied in reverse; the image encodes the dynamics dictated by the Hamiltonian and thus photoelectron spectroscopy can be done by the analysis of the image.

Recently, K\"ubel~{\it et al.}~\cite{Kubel17}, demonstrated a streak camera that allows to temporally resolve strong field ionization induced by linearly polarized short pulses. They called their method sub-cycle tracing of ionization enabled by infrared, i.e., STIER. The set-up used in the quoted experiment is a kind of pump-and-probe one, i.e., a few-cycle, intense, linearly polarized pulse in near-visible spectral range (VIS-pulse) induces ionization, whereas a moderately intense, mid-infrared pulse (IR-pulse) streaks photo-electrons allowing observation of the sub-cycle dynamics of strong field ionization. In particular, K\"ubel~{\it et al.} observed an asymmetry in the yield of low-energy electrons associated with the rescattering process. Interestingly, the respective asymmetry in the yield has not been fully reproduced by the authors through solving a one-dimensional (1D) TDSE, in spite of taking into account both the focal volume averaging (FVA) and integration over the Gouy phase. The discrepancy between experimental and computational results is ascribed to the observation that the 1D model cannot capture the details of the recollision processes accurately. Thus, the part of yield related to recolliding electrons is poorly modeled.

Here, we study strong field ionization in two-color fields corresponding to the STIER experiment. We solve the 2D TDSE within SAE approximation and collect data sets of intensities, phases, and delays significant enough to allow for the focal volume averaging and the integration over Gouy phase in a similar manner as in \cite{Kubel17}. We present calculated 2D momentum distributions and directly compare them to previously unpublished experimental data. The obtained 2D momenta distributions display complex features which we also analyze within the SFA framework. Our results reveal a complex ring structure that arises from the interference of attosecond wave-packets produced at different half-cycle maxima. The role of multiple electron rescattering in the formation of the asymmetry feature described in Ref.~\cite{Kubel17} is confirmed.

\section{Methods}
\subsection{TDSE simulation}
\begin{figure}[h]
    \centering
    \includegraphics[width= 0.95\columnwidth]{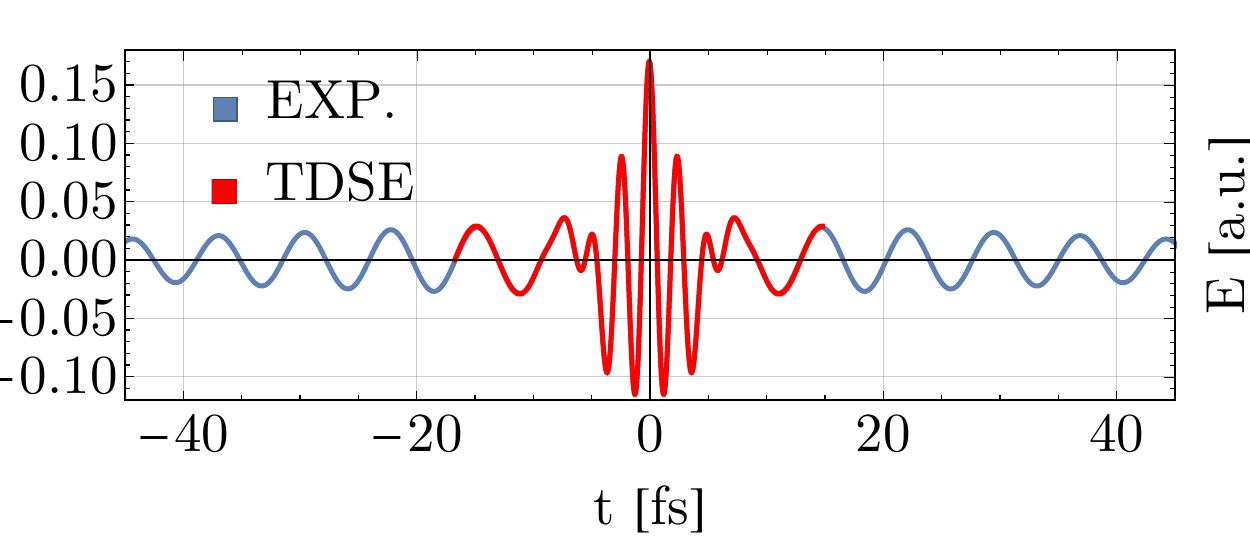}
    \caption{An exemplary pulse shape of IR + VIS field used for simulations (red lines, corresponding to $\phi=0$, $\tau=0[\tau_{\mathrm{IR}}]$) on top of part the experimental pulse (blue lines, the long decaying tails have been cut out from this picture).}
    \label{shortSignal}
\end{figure}

The main part of the calculations concerned with momenta distributions followed closely the method outlined in \cite{Kubel17}, but with upgraded dimensionality. Momenta distributions were found using 1D and 2D TDSE simulation performed in cartesian coordinates and length gauge which similarly assume a simplified description of the external field (see Fig.~\ref{shortSignal}). TDSE solver is based on the split operator method and the Fast Fourier Transformation algorithm as implemented in software developed by us for other restricted dimensionality models~\cite{Prauzner2007,Prauzner08,Efimov18,Thiede18}. The values of parameters used were $dt=0.05$ (evolution time step), $dx=\frac{100}{512}\approx 0.2$ (grid spacing), $N=28672$ (number of grid point in one direction). The groundstate was found using propagation in imaginary time. Both pulses were linearly polarized with respect to $z$ axis and the respective wavelengths and peak intensities were: ($\lambda_\mathrm{VIS} = 735$ nm, $I_\mathrm{VIS} = 7\times 10^{14}$ W cm$^{-2}$)  and  ($\lambda_\mathrm{IR}=2215$ nm, $I_\mathrm{IR}=3\times 10^{13}$ W cm$^{-2}$).
The pulse shape has been identical to the one from the original paper of K\"ubel {\it et al.}~\cite{Kubel17}, namely: the IR-pulse is approximated by a $cos$ carrier wave devoid of envelope, with a significantly shorter duration (as compared with experiment) of 4.25 cycles of length $\tau_\mathrm{IR} \approx 7.4$ fs each and 0 field strength at the beginning. The envelope for VIS-pulse has a full width at half maximum (FWHM) of 5 fs (see Fig.~\ref{shortSignal}):
\begin{equation}
\bm{E}(t) = E_\mathrm{VIS}(t)\bm{e}_z+E_\mathrm{IR}(t)\bm{e}_z
\end{equation}
where
\begin{eqnarray}
\label{def:fields}
E_\mathrm{VIS}(t) &=& F_\mathrm{VIS}f(t-\tau)\cos(\omega_\mathrm{VIS}(t-\tau)+\phi),\\
E_\mathrm{IR}(t) &=& F_\mathrm{IR}\cos(\omega_\mathrm{IR}t).
\end{eqnarray}
Here $F_i$ and $\omega_i$ stand for field amplitude and frequency of corresponding pulse, respectively, $f(t-\tau)$ is a Guassian envelope of VIS pulse, $\tau$ is the delay between pulses and $\phi$ is the CEP of the VIS signal (in the later part of this article we will use the symbol $\phi$ and CEP interchangeably).

\subsection{Post-processing and fitting data to experiment}

The well-known affliction of laboratory strong field experiments is the inherent spatial averaging due to non-uniformity of laser beam cross-section and geometry of the optical elements.

The most straightforward recipe for adapting theoretical calculations is to average over intensities with the weight of inverse intensity \cite{strohaber2015highly}, in other words, calculating the integral \begin{equation}
    \label{vol-avg}
    \int_0^{I_0} dI P(I)/I.
\end{equation} 
On the other hand, more advanced measures could be employed taking into account the Gouy phase \cite{paulus2005measurement}. 
Surprisingly, in our case these averaging methods did not lead to improvement over the non-averaged (single intensity) results, to the contrary - their use (especially accounting for Gouy phase) caused an underestimation in the width of momentum spectra when a single delay was considered.
This can be understood, as follows. 
In practice, the action of FVA can be reduced to mixing the thin (in terms of width) low intensity momentum distributions with the wide high intensity momentum distributions. Since the former usually obtain much higher weights, the averaged distribution appears thinner than the original one.
On the other side, in the experiment, the synchronization jitter between IR and VIS pulses and the focal geometry can effectively lead to uncertainties in the time delay which broadens the distributions.
Moreover, FVA becomes harder to implement correctly when two-color pulses are considered.
Consequently, we have restricted our analysis to the simple volume average recipe given by eq.~\ref{vol-avg} (FVA), delay ($\pm$0.8 fs) average (DA) or both (DFVA).

\begin{figure}[ht]
    \centering
    \includegraphics[width= 0.95\columnwidth]{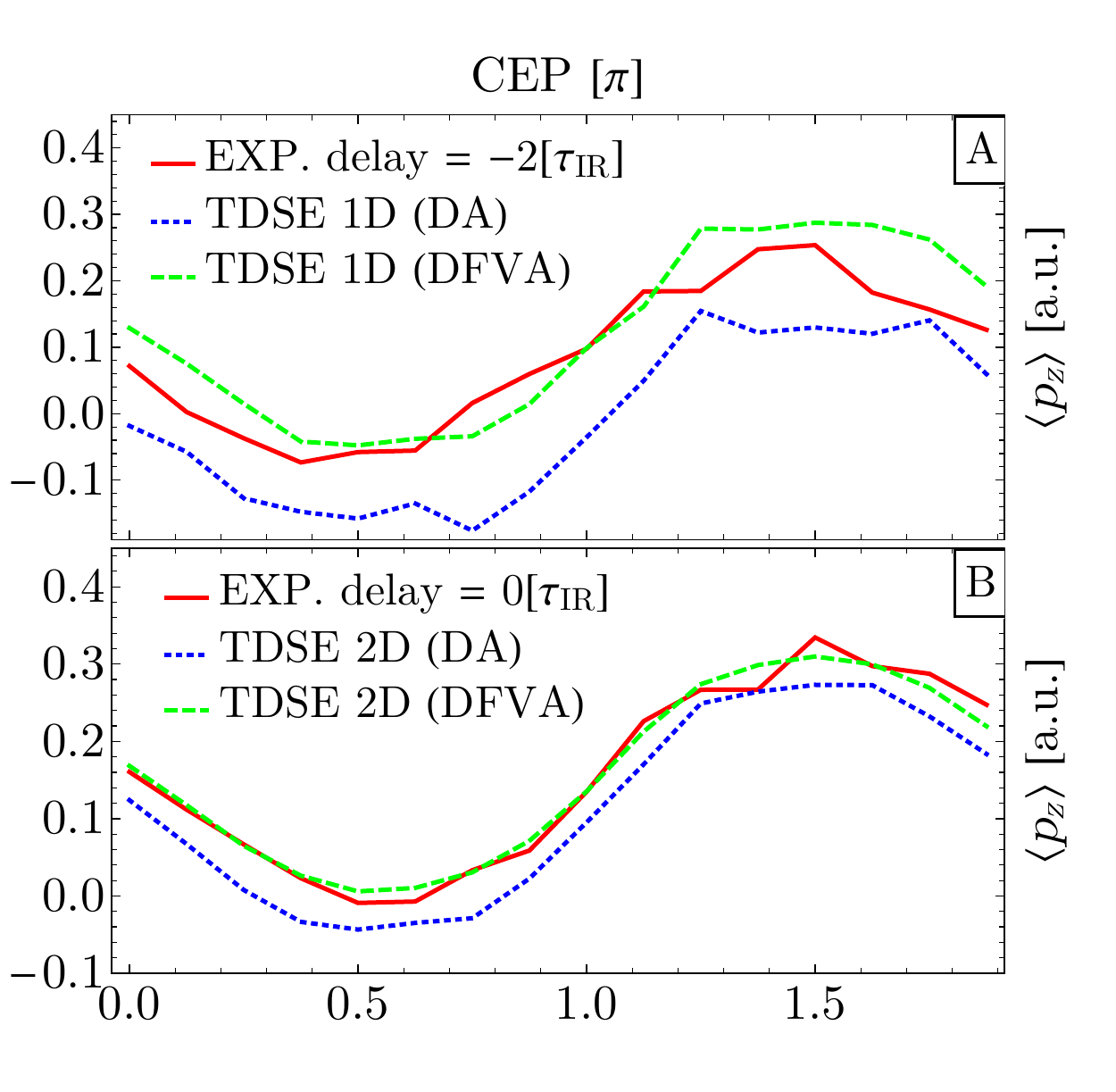}
    \caption{Comparison of mean $p_z$ value of the momentum distribution ($\langle p_z \rangle$) dependence on CEP between experiment (red, solid) and 1D (panel A) or 2D (panel B) simulation (dashed and dotted lines). Notice that even simulations fit to the same physical moment separated by 2 IR periods ($\approx 14.8$ fs).}
    \label{moments}
\end{figure}

After averaging the obtained momentum distributions were smoothed out using a median filter \footnote{The non-averaged data was smoothed with median filter of 10, whereas the data that has been focal-volume averaged (FVA) were smoothed with median filter of size 1 (the highly oscillatory behavior has been mostly canceled thanks to FVA averaging).} and (point) resampled to the (smaller) resolution of experimental results.
Since the experiment provided data over a wide range of delay, but the absolute information about the CEP ($\phi$ of eq. \ref{def:fields}) was not determined completely, the CEP dependence of TDSE and experimental results had to be compared and matched.
To aid our analysis we've employed a secondary measure of (dis)similarity, based on the least squares method for partially smoothed out and normalized central momenta distributions. This automated method yielded the CEP shift and the experimental delay $0[\tau_\mathrm{IR}]$ (corresponding to the center of the VIS pulse centered at a maximum of the IR field, see Fig.~\ref{shortSignal}, also delay $\approx 5.87$ fs from \cite{Kubel17}), as the most consistent with our 2D results (Fig.~\ref{moments} B) for raw data and DFVA data. In case of 1D results (Fig.~\ref{moments} A), the raw CEP spectrum matched to delay $-2[\tau_\mathrm{IR}]$ (separated by 2 $\tau_\mathrm{IR}$ from the 2D match), but the DFVA matched with the reversely scattered electrons (incorrect). This result suggests that automated matching experimental data to low dimensional models can be deceitful 
and some additional information is needed for a successful match. Nevertheless, such problems become absent in the case of higher dimensional models as Fig.~\ref{moments} B suggests.

\section{Results}
\subsection{1D momenta distributions}
\begin{figure}[h]
    \centering
    \includegraphics[width= 0.95\columnwidth]{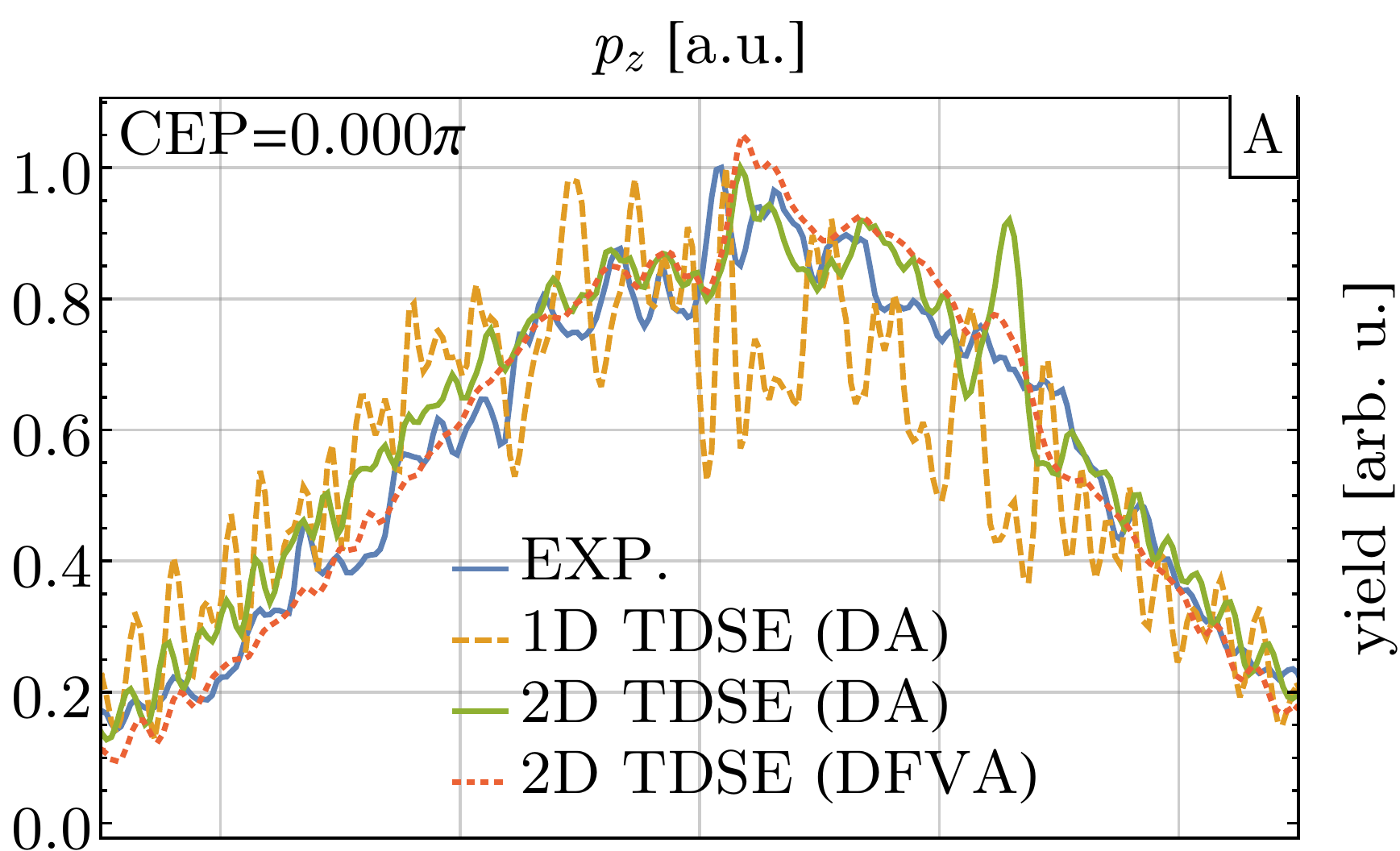}
    \includegraphics[width= 0.95\columnwidth]{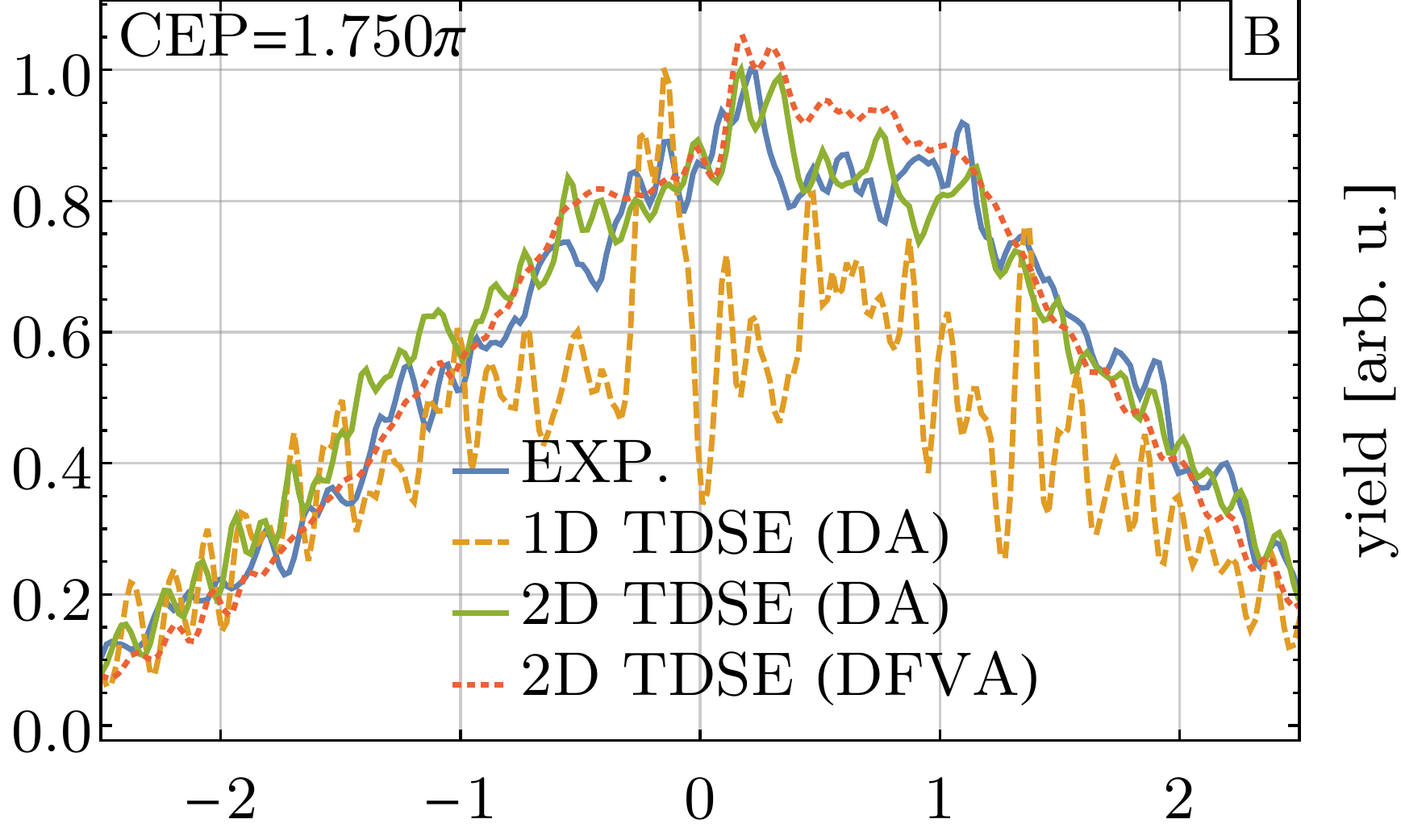}
    \caption{Comparison of momenta distributions for $\tau=0[\tau_{\mathrm{IR}}]$, at different CEP values for experimental data (blue, solid), nearby-delay averaged (DA) 1D sim. (yellow), 2D simulation (green, solid) and 2D with focal-volume and delay ($\pm$ 0.8 fs) averaging (DFVA).}
    \label{comp0}
\end{figure}

\begin{figure}[h!]
    \centering
    \includegraphics[width=1.0\columnwidth]{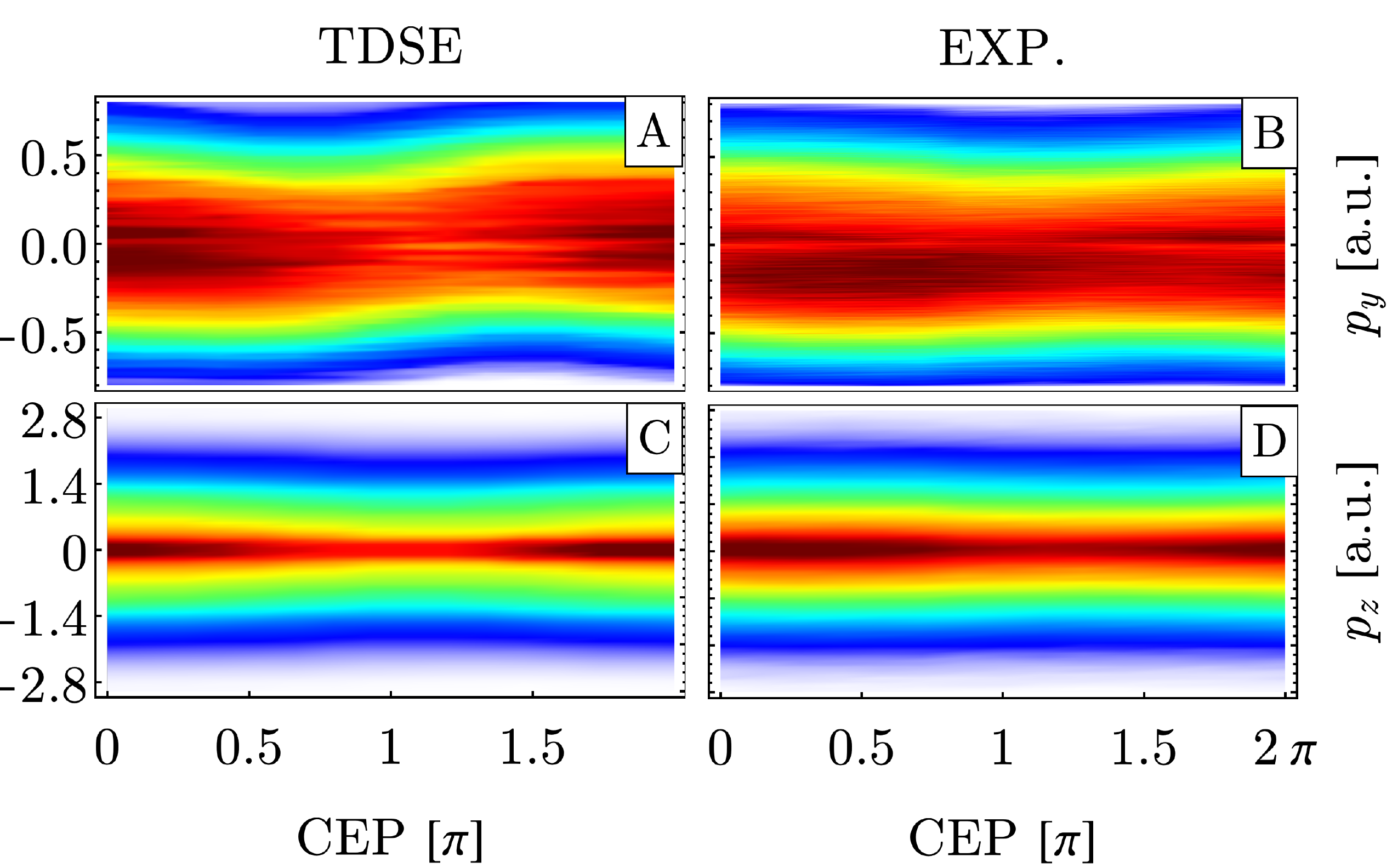}
    \caption{Comparison of non-averaged, smoothed, single delay (0$[\tau_\mathrm{IR}]$) momenta distributions for both axes obtained in simulation (A, C - left column) and experiment (B, D - right column) for different CEP values}
    \label{CEPpzpy}
\end{figure}

Performing a full comparison of 1D and 2D TDSE results at $\tau=0[\tau_\mathrm{IR}]$ (see Fig.~\ref{shortSignal}) with experiment, we notice (see Fig.~\ref{comp0} and compare with Fig.~4c of \cite{Kubel17}) good agreement at most CEP's for solely delay averaged ($\pm 0.8$ fs), otherwise raw data (DA) as well as for delay and focal-volume averaged (DFVA) data. 
Advantages of one average versus the other, seen previously on Fig.~\ref{moments}, are not evident at this point; however, the improvement over the 1D result is visible. 
Indeed, the agreement is reasonable for both projected directions even without FVA over a range of CEP values and fixed delay, see Fig~\ref{CEPpzpy}.
Remaining discrepancies may be attributed to low experimental resolution leading to significant uncertainties in fitting parameters and a limited number (3) of averaged delays.

\subsection{2D momenta distributions}

\begin{figure}[ht!]
    \centering
    \includegraphics[width=0.99\columnwidth]{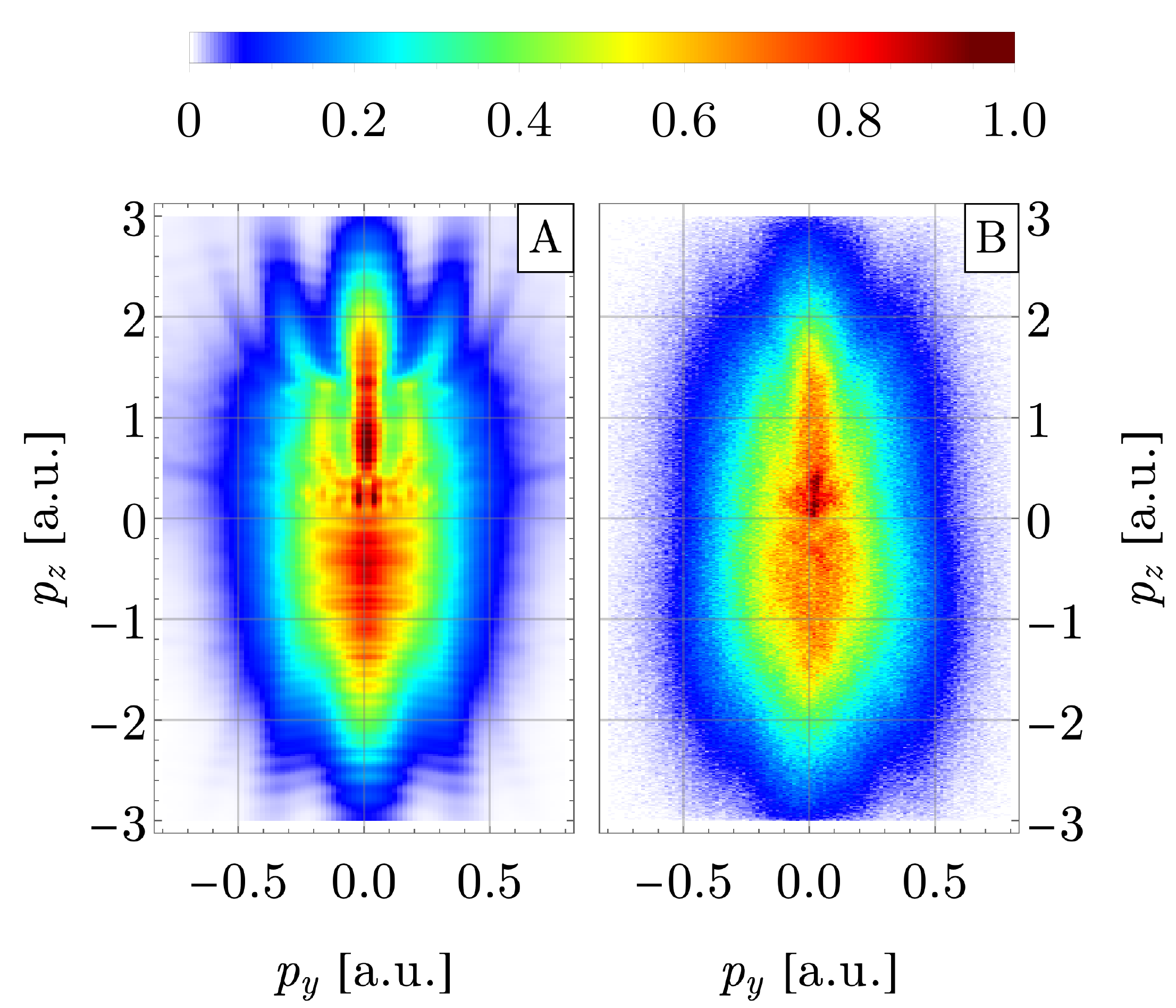}
    \caption{CEP averaged and smoothed 2D photoelectron momentum distributions for $\tau=0[\tau_{\mathrm{IR}}]$ parallel ($p_z$) and perpendicular ($p_y$) to the laser polarization direction for TDSE (A) compared with experiment (B).}
    \label{comp2d}
\end{figure}

\begin{figure*}[h!t]
    \centering
    \includegraphics[width=2.0\columnwidth]{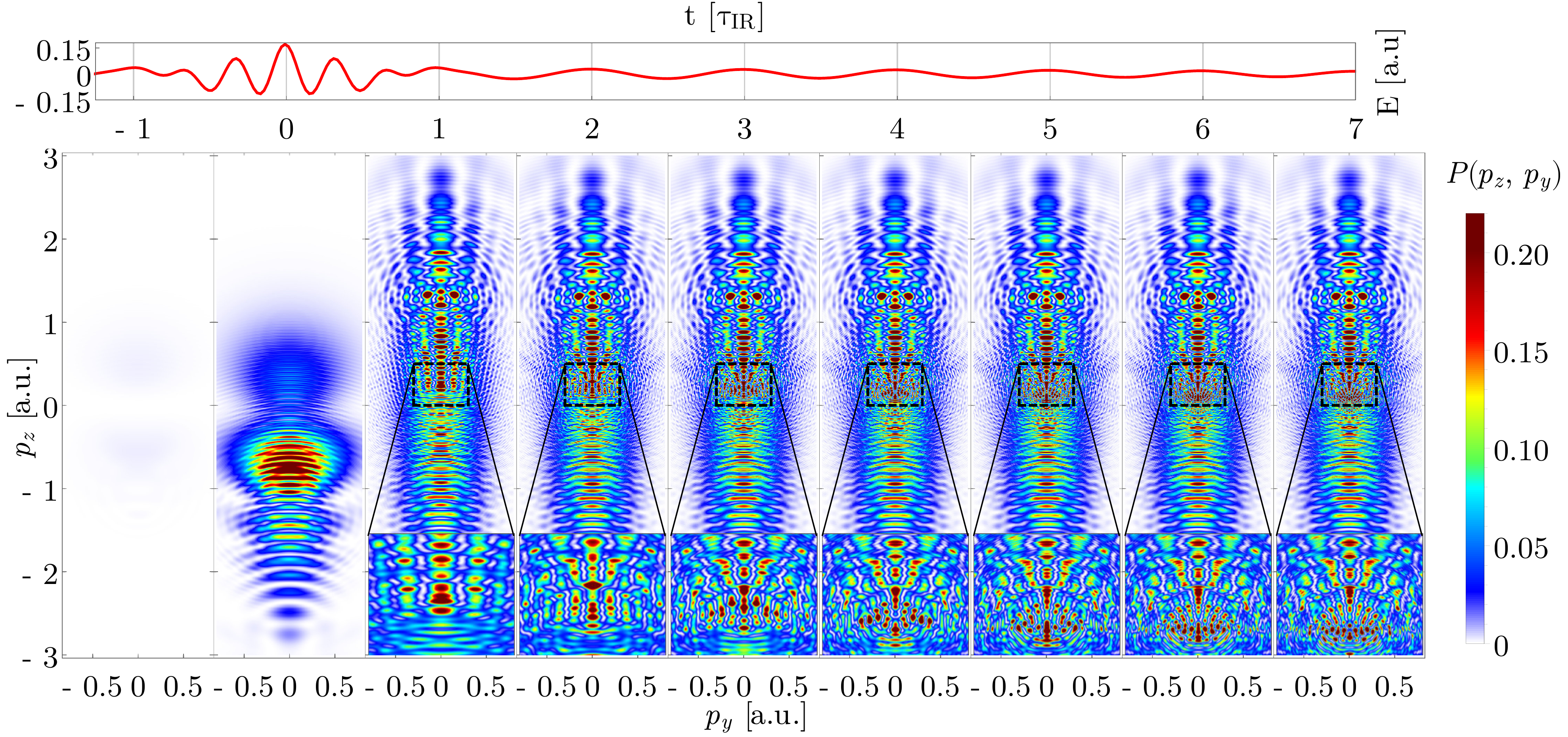}
    \caption{Momentum distributions snapshots at equal IR period intervals for exemplary $delay=-1 [\tau_\mathrm{IR}]$, $CEP=0$. The overall distribution appears be stable after the second cycle ($\tau_{\mathrm{IR}}$), significant changes appearing in the LES area (zoom on the area $p_z = (0,0.5), p_y=(-0.3,0.3)$ presented).}
    \label{ev2d}
\end{figure*}

Comparison of CEP averaged photoelectron momentum maps predicted by theory (panel A) versus experiment (panel B) is presented in Fig.~\ref{comp2d}. Apart from good agreement between the two results (the theoretical prediction has been smoothed out to aid comparison), we observe large asymmetry on the $p_z$ axis. The upper side ($p_z>0$) is much narrower in $p_y$ than the lower side and exhibits the asymmetric feature at $p_z = 0.2$ a.u., which emerges due to multiple electron rescattering events and Coulombic interaction \cite{Kubel17}. 
In order to confirm this statement, in Fig.~\ref{ev2d} we present snapshots of momentum photoelectron distributions (for fixed CEP and delay value)
at time instances separated by multiple IR periods (corresponding to zeroes of vector potential \cite{becker2018plateau}). The obtained structures remain mostly fixed after the high intensity VIS pulse is over (see times $t>1 [\tau_{\mathrm{IR}}]$ in Fig.~\ref{ev2d}), with the exception of low energy regions for which (for convenience) we provide zoomed windows. Indeed, one can observe the formation of a holographic, shell-like structure concentrated around $p_z = 0.2$. The asymmetry builds up with each IR period through simultaneous horizontal ($p_y$) splitting and shifting towards lower $p_z$ values.
This non-equilibrium steady structure emerges through interference of both forward and backward rescattered electrons accelerated by the IR field \cite{peng2015tracing, geng2014attosecond}.
Comparing time $t=1 [\tau_{\mathrm{IR}}]$ of Fig.~\ref{ev2d} with times $t>1 [\tau_{\mathrm{IR}}]$ we see a significant increase in phase accumulation in the radial direction from the $(p_y=0,p_z=0.2)$, leading to sharp drops in intensity visible in the zoomed regions, the effect again attributable to multiple rescatterings.
The significance of this asymmetry is expected to grow with the wavelength of the IR field \cite{geng2014attosecond}.
\begin{figure}[h!t]
    \centering
    \includegraphics[width=0.99\columnwidth]{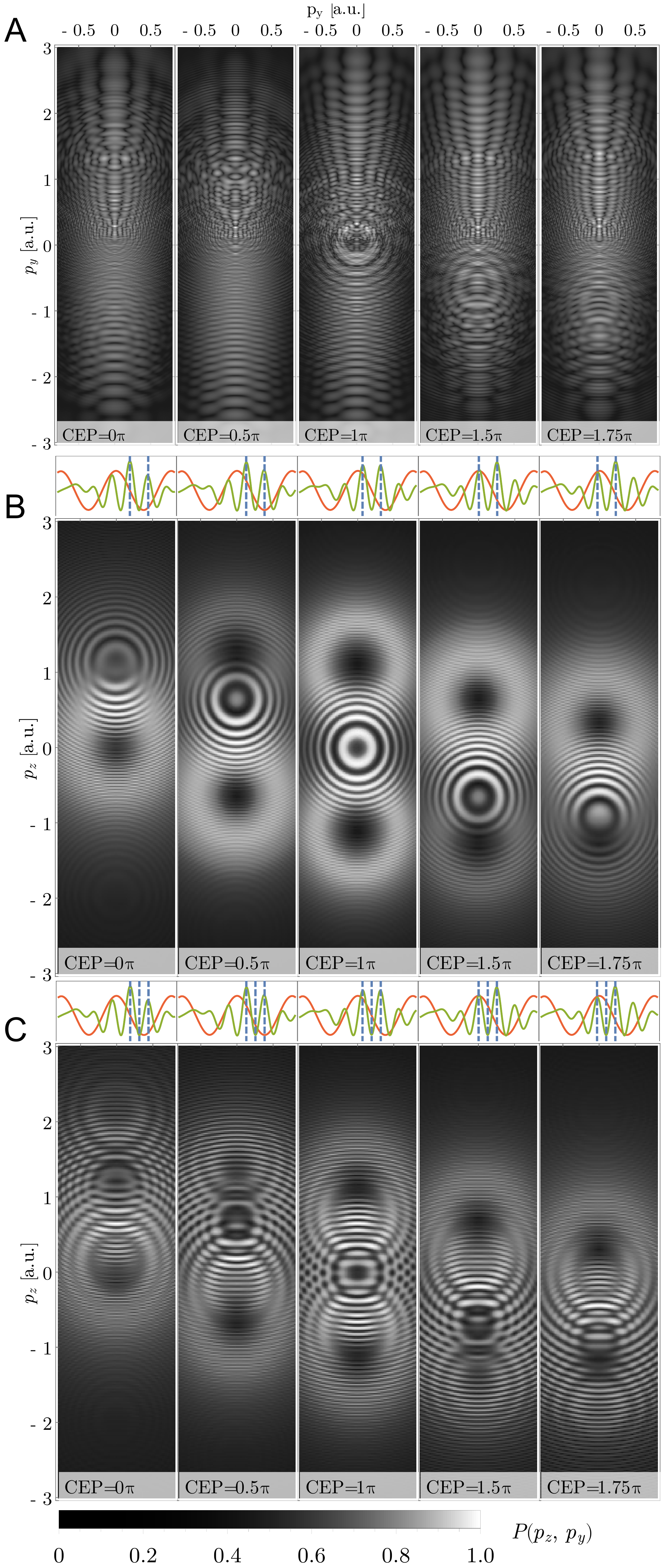}
    \caption{Photoelectron momentum maps for individual CEP's. A) TDSE result, B) SFA prediction for ionization times $\{t_0, t_2\}$, C) SFA prediction for ionization times $\{t_0, t_1, t_2\}$ (details in text). The diagrams in B and C present IR vector potential (red solid line), total electric field (green solid line), ionization times (dashed blue lines).}
    \label{ringEV}
\end{figure}

Returning to Fig.~\ref{comp2d}; at $p_z>0$ and around $p_y = \pm~0.2$~a.u. in both, experiment and simulation, one can notice sidebands parallel to the polarization axis, also known as holographic structures \cite{huismans2011timea}. The feature is visibly less pronounced in the experiment than in the simulation due to limited resolution (high-resolution experiments using the STIER technique are currently in preparation).
In general holographic structures can be attributed to interference of two coherent wave packets ionized at two nearby (same quarter-cycle) instances of time. One of such wave packets (signal) is assumed to rescatter with the parent ion before meeting the other (reference) wave packet. These kind of processes occur frequently in this setup as can be seen on the positive part of the $p_z$ axis in Fig.~\ref{ringEV} A, which presents photoelectron momentum maps in function of VIS field CEP ($\phi$) value.
The frequent occurrence of rescattering events is also the reason why such side lobes are visible on CEP averaged plots such as Fig.~\ref{comp2d}.

\begin{figure}[h!t]
    \centering
    \includegraphics[width=.99\columnwidth]{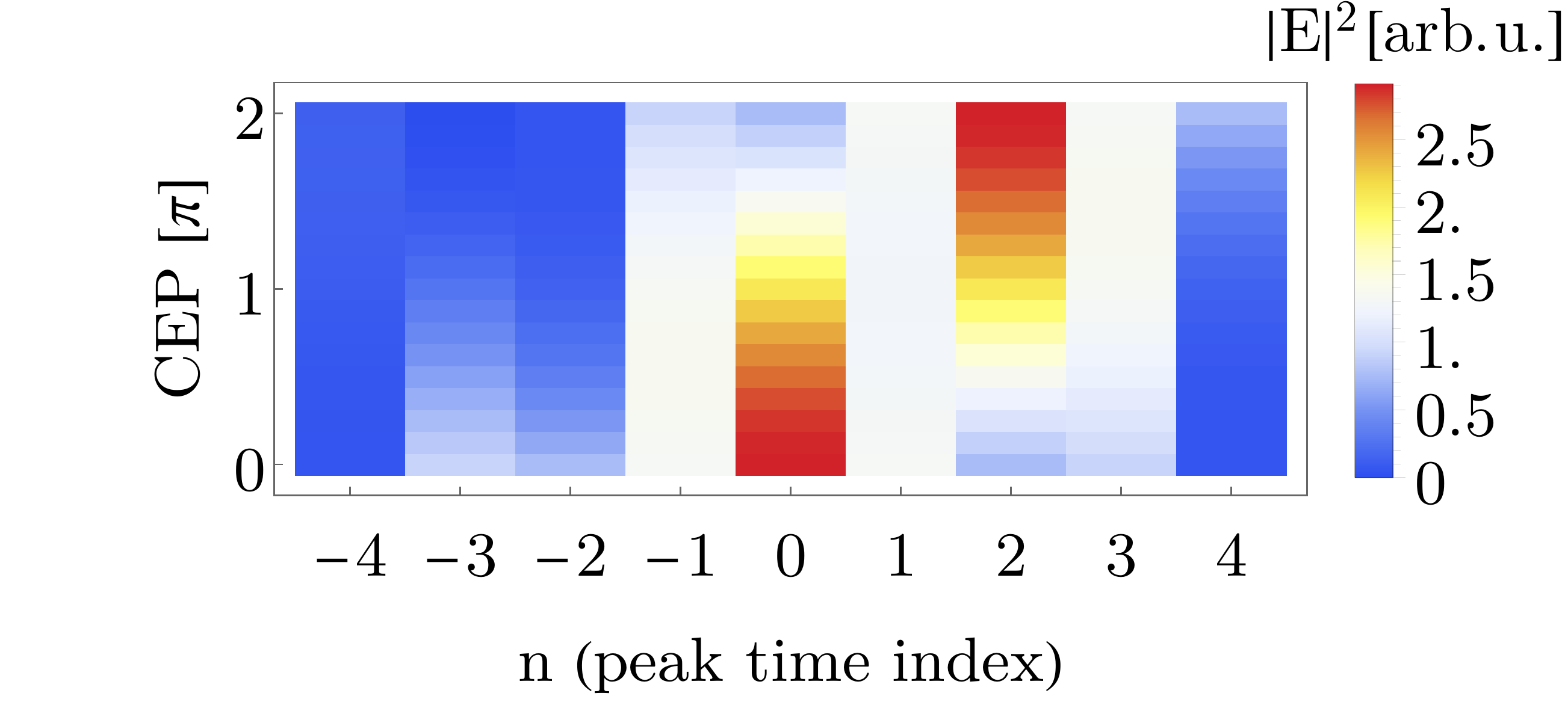}
    \caption{Field intensity at ionization times $t_n$ for different CEP values and $\tau=0[\tau_{\mathrm{IR}}]$. The dominant peaks throughout the whole CEP domain are $\{t_0,t_2,t_1\}$.}
    \label{heat_cep}
\end{figure}

The holographic structures seen on the detailed maps of Fig.~\ref{ringEV}~A are also heavily influenced by other sub-cycle effects, including interference of wave packets ionized with half, one or more cycle delays. One of the most dominant traces of this is the ring structure centered at $p_z=0$ a.u. for CEP=$\pi$ and of approximately $0.25$ a.u. (first) radius (see Fig.~\ref{ringEV}~A, third panel from the left). This ring structure is displaced along the $p_z$ axis with the change of CEP value (see Fig.~\ref{ringEV}~A). In experiments using the RABBITT technique the amount of ring displacement was shown \cite{remetter2006attosecond, varju2006angularly} to be proportional to the area under the vector potential between two dominant ionization times in some simpler models. 
Arguably, the presented situation is considerably more complex than the setups occurring in the cited references. In particular, the vector potential is the sum of IR and VIS contributions, i.e. $\bm{A}(t)=\bm{A}_{\mathrm{IR}}(t)+\bm{A}_{\mathrm{VIS}}(t)$, and therefore does not behave as a simple sine-like function. Nevertheless, given the versatility of the SFA \cite{amini2018symphony} we suspect that a similar analysis might be valuable. Thus, resorting to SFA analysis we show that the CEP displacement of the ring is in fact due to interference between few~(2~or~3) major ionization times (see Fig.~\ref{heat_cep}) and only weakly dependent on the $\bm{A}_{\mathrm{VIS}}(t)$ contribution. 

\subsection{SFA analysis}
The ionization amplitude $a_p$ and action $S(t)$ are given by:
\begin{eqnarray}
    \label{ap}
    a_p &=& -i \int_{-\infty}^{\infty}\bm{E}(t')\cdot \bm{d}(\bm{p}+\bm{A}(t')) e^{-i S(t')} dt', \\
    S(t)&=&\int_t^\infty \left[ \frac{(\bm{p}+\bm{A}(t'))^2}{2} + I_p\right] dt',
    \label{actionS}\\
\bm{d}(\bm{p})&=&\frac{\bm{p}}{(\bm{p}^2+2 I_p)^3},
\end{eqnarray}
where $I_p$ is the ionization potential.
Assuming that the ionization takes place at the extrema of the total electric field, which in our case can be approximated with the extrema of the VIS field, we get:

\begin{equation}
t_n = (n-\frac{\phi}{\pi})\tau_\mathrm{VIS}/2,
\label{timesCEP}
\end{equation}
for most likely times $t_n$ of ionization.
Using this, equations (\ref{ap}-\ref{actionS}) can then be reduced to:
\begin{align}
    \label{ap2}
    a_p ={}& \sum_{n} E_z(t_n)d_z(\bm{p}-\bm{A}(t_n))e^{-i S(t_n)},\\
    \begin{split}
        S(t_n) ={}& -(\frac{\bm{p}^2}{2}+I_p)t_n\\ 
        &-p_z \int_0^{t_n} \left[ A(t') + \frac{A(t')^2}{2} \right] dt'.
        \label{actionS2}
    \end{split}
\end{align}
The resulting probability $|a_p|^2$ is presented as a function of CEP in Figs.~\ref{ringEV} B, C.
For the sake of clarity, above each momentum distribution in Fig.~\ref{ringEV} B and C, we present a diagram with VIS+IR pulse (green lines) and the IR vector potential in the background (red lines). Blue dashed lines point to ionization times taken into account in the analysis.
\begin{figure}[h!t]
    \centering
    \includegraphics[width=0.99\columnwidth]{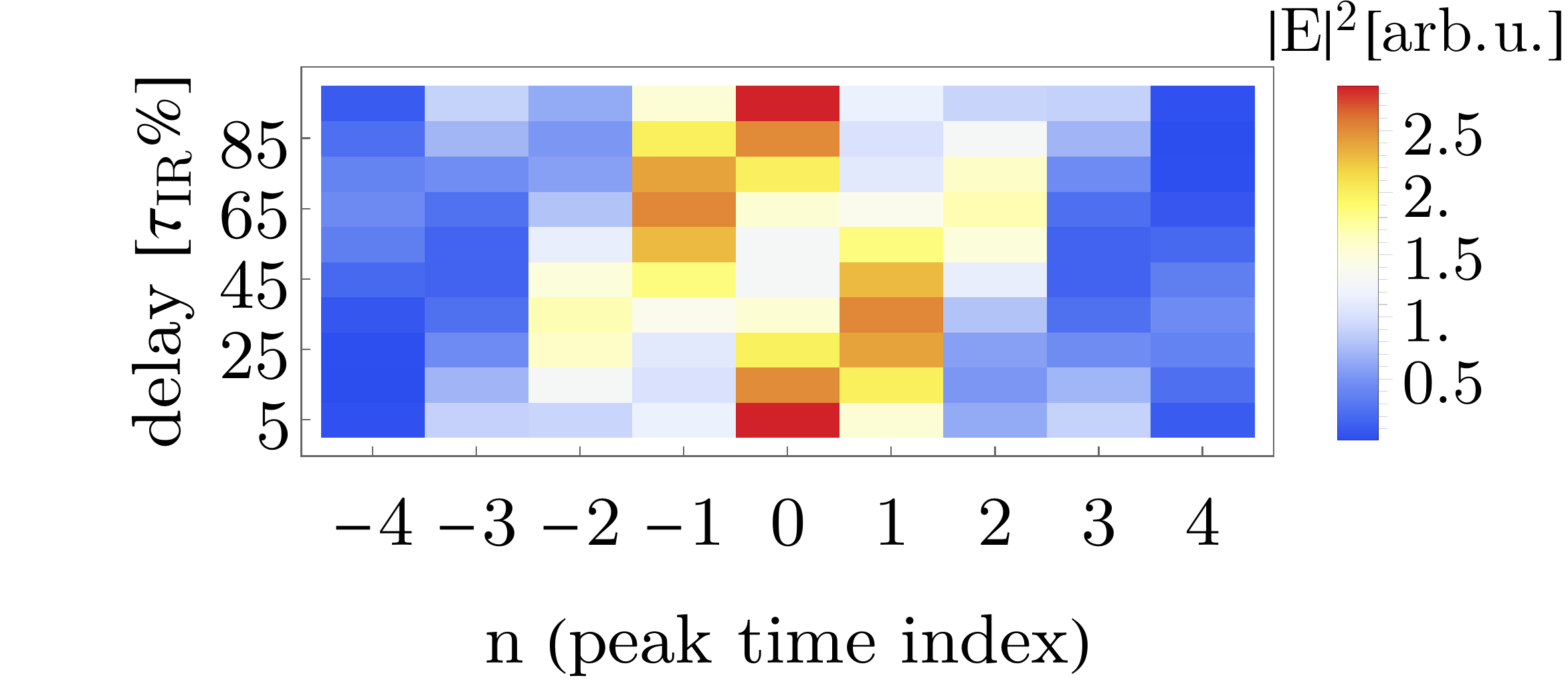}
    \caption{Field intensity at ionization times $t_n$ for different delay values and CEP=0. The set of dominant peaks varies with the delay, leading to less straightforward analysis.}
    \label{heat_delay}
\end{figure}
\begin{figure*}[h!t]
    \centering
    \includegraphics[width=0.92 \textwidth]{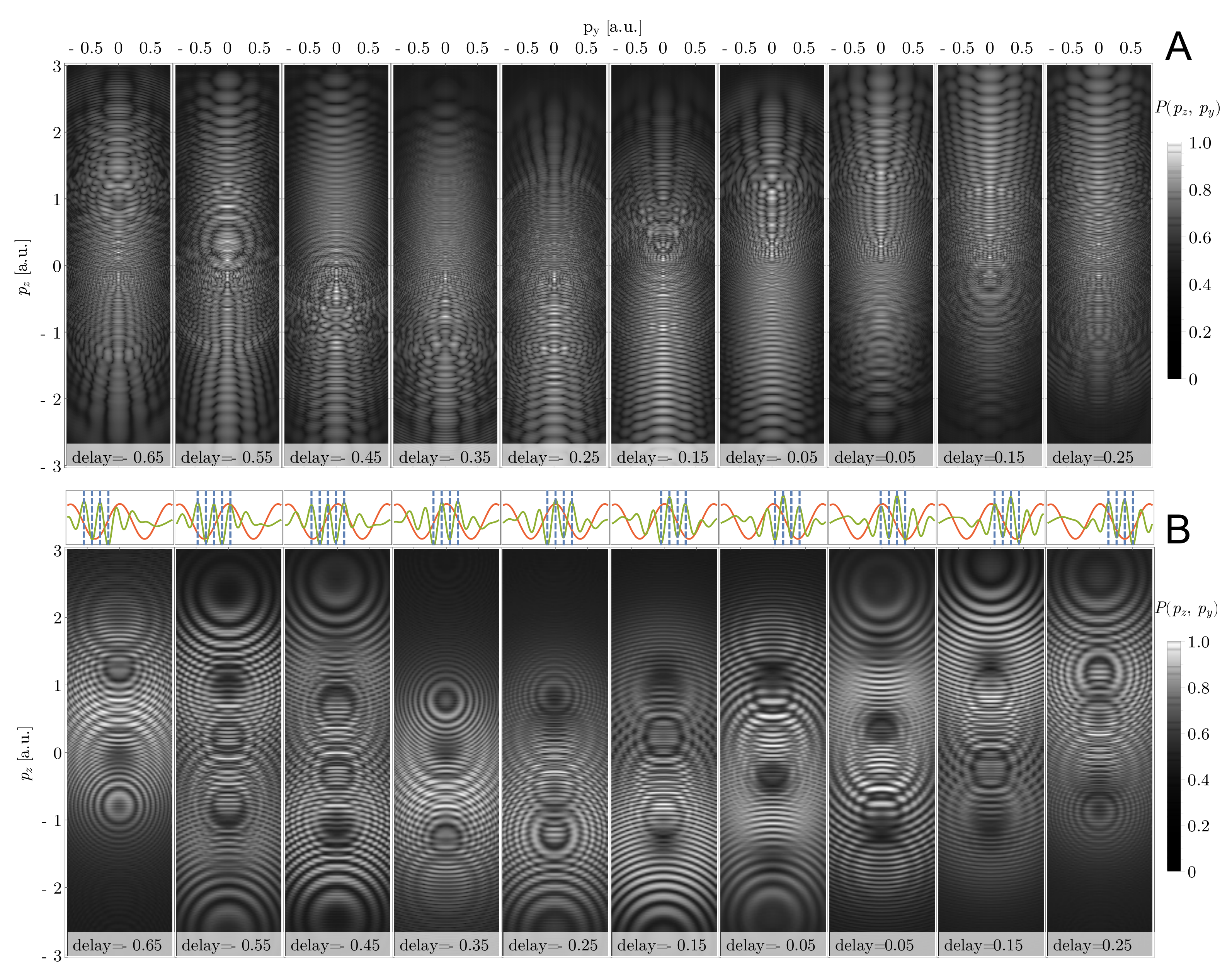}
    \caption{Photoelectron momentum maps for individual delays (CEP=0). A) TDSE result, B) SFA prediction for ionization times at which electric field surpasses a given threshold  (details in text). The diagrams in B and C present IR vector potential (red solid line), total electric field (green solid line), ionization times (dashed blue lines).}
    \label{delays}
\end{figure*}
First, we take into account two ionization times $\{t_0,t_2\}$ (Fig.~\ref{ringEV}~B), for which the field points in the same direction. 
At CEP=0 one can notice the ring structure centered at just above $p_z=1$, which intensifies while moving down towards lower $p_z$ values with the increasing value of CEP. 
For CEP=$\pi$ the contributions from ionization times $\{t_0,t_2\}$ become equal (see Fig.~\ref{heat_cep}) leading to a pronounced ring, centered at $p_z=0$ with the first radius of approx. 0.25 a.u., in agreement with TDSE results. 
The rings are preserved even when the IR vector potential alone is considered ($\bm{A}_{\mathrm{VIS}}(t)$ neglected) in calculating $a_p$ with respect to eqs.~(\ref{ap2}-\ref{actionS2}) (results not shown). Moreover, the contribution of the term $A(t')^2/2$ has small influence on the ring and can be dropped in the eq.~(\ref{actionS2}). 
Thus, for readability in diagrams above each momentum distribution, in Fig.~\ref{ringEV} B and C, only $A_{\mathrm{IR}}$ is shown for reference (red lines). 

In Fig.~\ref{ringEV}~C we expand the analysis to the case of three ionization times, $\{t_0, t_1, t_2\}$. The interference structures become richer and the ring acquires slight deformations. The addition of more ionization times, does not bring clarity to the overall picture. Other effects, such as multiple rescatterings with the parent ion, provide additional structures which suppress or enhance the visibility of the ring and its surrounding features. 

In case of changing delay instead of the CEP value, the eq.~(\ref{timesCEP}) changes to:
\begin{equation}
    t_n = n\tau_\mathrm{VIS}/2 + \tau,
\end{equation}
while the eqs.~(\ref{ap2}-\ref{actionS2}) stay unchanged. 
Now, in contrast with the varying CEP case (see Fig.~\ref{heat_delay} and compare it with Fig.~\ref{heat_cep}) it is not trivial to select the ionization peaks contributing the most to the momentum maps pictured in Fig.~\ref{delays}~A. In fact, many such choices exist. Here we limit our analysis to ionization peaks for which the electric field intensity passes a chosen threshold ($|E(t_n)|^2>0.009$ a.u.) and present the results in Fig.~\ref{delays}~B. The abundance of ring-like features could be discouraging at first, but through careful inspection one can see that the vast majority of the ring like structures of Fig.~\ref{delays}~B either, can be found directly (as rings) or appear to steer the holographic structures found in the TDSE result (Fig.~\ref{delays}~A). Although, our analysis is by no means exhausting we have shown how to understand one layer of complexity of the TDSE result and how other interference structures can be affected. 

\section{Discussion and summary}

Through an increase in dimensionality of the TDSE simulation we have found that individual (per CEP) 1D momenta distribution can indeed be substantially improved.
However, fitting computational results to experimental ones requires caution - especially when comparing reduced dimensionality models to experimental data. Averaging over focal volume and Gouy phase needs to be done on par with averaging over "nearby" delays in order to retain the width of the momenta distributions. 
On the other hand, the individual 2D momenta distributions could not be directly compared due to limited resolution of the experimental results.
Using the CEP averaged results we have found good agreement between theory and experiment. Asymmetric features present in the experimental results have been traced to multiple rescattering effects in the IR field; visible and similar at all CEP values.
The complex individual momentum maps consist of two kind of structures: holographic lobes extending in the direction perpendicular to the polarization direction, ATI structures (affecting the holographies) and/or ring structures. In particular, an intense ring centered at CEP=$\pi$ and traversing through the $p_z$ axis with the change of CEP value have been noted. 
With the use of the SFA we have shown that this ring originates from the interference of two or more wave packets born at different ionization times. In a bit less straightforward way this effect can also be seen when delay instead of CEP is manipulated.
Together with holography, the ring structures record the electron dynamics on a sublaser-cycle time scale that lead to applications in next generation photoelectron spectroscopy.
Complementary analysis from the perspective of high harmonic generation in the given setup is currently underway.

\bigbreak
\section*{Funding Information}
This project has received funding from the EU’s Horizon2020 research and innovation programme under the Marie Sklodowska-Curie Grant Agreement No. 657544 (MK);\\
National Science Centre, Poland via project No. 2016/20/W/ST4/00314 (MM, JPB and JZ).\\
We also acknowledge the support of PL-Grid Infrastructure.
\bigbreak
\bigbreak
\bibliography{reflist,reflistMM}



\end{document}